\documentstyle[preprint,tighten,aps,epsf]{revtex}
\begin{document}
\preprint{MKPH-T-96-20}
\draft
\title{Generalized polarizabilities and the spin-averaged amplitude in
virtual Compton scattering off the nucleon}
\author{D.\ Drechsel, G.\ Kn\"ochlein, A.\ Metz, S.\ Scherer}
\address{Institut f\"ur Kernphysik, Johannes Gutenberg-Universit\"at,
J.\ J.\ Becher-Weg 45, D-55099 Mainz, Germany}
\date{August 28, 1996}
\maketitle
\begin{abstract}
   We discuss the low-energy behavior of the spin-averaged amplitude of
virtual Compton scattering (VCS) off a nucleon.
   Based on gauge invariance, Lorentz invariance and the discrete symmetries,
it is shown that to first order in the frequency of the final
real photon only two generalized polarizabilities appear.
   Different low-energy expansion schemes are discussed and put into
perspective.
\end{abstract}

\section{Introduction}
   Virtual Compton scattering (VCS) off the proton, as tested in, e.g.,
the reaction $e^-+p\to e^-+p+\gamma$, has recently attracted considerable
interest \cite{VCSproc}.
   Several experiments have been proposed
\cite{Audit,Brand,Breton,Audit2,Shaw},
utilizing the opportunities which a virtual spacelike photon offers, namely,
an additional longitudinal polarization degree of freedom and the fact that
energy and momentum transfer of the virtual photon can be varied independently.
   At the same time, in comparison with real Compton scattering, the extraction
of new experimental information will be more difficult since the process
$e^-+p\to e^-+p+\gamma$ contains an interference between VCS and the 
Bethe-Heitler contribution, describing radiation off the electron.
   On the theoretical side, the low-energy theorem (LET) of Low \cite{Low}
and Gell-Mann and Goldberger \cite{GellMann}
has lately been extended to include virtual photons as well
\cite{Guichon,Scherer}.
   The structure-dependent part beyond the LET was analyzed in \cite{Guichon}
in terms of a multipole expansion.
   Keeping only terms linear in the energy of the final real photon,
the model-dependent amplitude was parametrized in terms of ten
``generalized polarizabilities'', and these polarizabilities were evaluated
in the framework of a nonrelativistic quark model \cite{Guichon,Liu}.
   Predictions for the spin-averaged polarizabilities $\alpha(|\vec{q}|)$
and $\beta(|\vec{q}|)$ were obtained by several authors within various
frameworks, such as an effective Lagrangian including resonances and
$t$-channel exchanges \cite{Vanderhaeghen}, the linear $\sigma$ model
\cite{Metz}, and the heavy-baryon formulation of chiral perturbation theory
\cite{Hemmert}.
   An alternative low-energy expansion for virtual Compton scattering off a
spin-zero target, and thus implicitly also for the spin-averaged part of the
nucleon VCS amplitude, has been obtained in \cite{Fearing}.

   In \cite{Metz} an interesting observation was made:
within the framework of the linear $\sigma$ model only two of the three
scalar generalized polarizabilities introduced in \cite{Guichon} were found
to be independent.
   In the following, we will reinvestigate the spin-independent part
of the VCS amplitude and demonstrate that the findings of \cite{Metz}
can be proven to be a general consequence of charge-conjugation
symmetry combined with crossing symmetry.
   Unless charge-conjugation invariance is violated, there are only two
independent scalar generalized polarizabilities.
   Furthermore, we illustrate how the standard limit of real Compton scattering
is naturally obtained, if the expansion in the final-photon energy is not
truncated at first order.
    Finally, the low-energy expansion of \cite{Fearing} and its application 
in the framework of a heavy-baryon calculation \cite{Hemmert} are 
connected to this work.

\section{General Formalism}   
   For the purpose of simplicity, we consider the VCS amplitude for a
spinless target, e.g., a positively charged pion: $\gamma^\ast(q,\epsilon)
+\pi^+(p)\to\gamma^\ast(q',\epsilon')+\pi^+(p')$.
   In the following discussion, we will thus always refer to the pion,
but the general results also apply to the spin-averaged
amplitude of VCS off the proton \cite{Tarrach,Bernabeu},
which is, of course, the reaction of current experimental and 
theoretical interest \cite{VCSproc}.

   Using the conventions of Bjorken and Drell \cite{Bjorken}, the invariant
amplitude may be written as
\begin{equation}
\label{m}
{\cal M}=-2Mie^2 \epsilon_{\mu} \epsilon_{\nu}'^\ast M^{\mu\nu},
\end{equation}
where $\epsilon$ and $\epsilon'$ denote ``polarization vectors'' of
the initial and final photon, respectively, 
$e>0$ is the elementary charge ($e^2/4\pi=1/137$),
and $M$ is the mass of the target,
here the pion.
   According to \cite{Bjorken}, the normalization of the respective 
invariant amplitude ${\cal M}$ differs by a factor of $2M$ between the pion
and proton case.
   When considering VCS off the proton, one therefore has to omit this factor 
in Eq.\ (\ref{m}) in order to obtain the same normalization of the Compton
tensor $M^{\mu\nu}$ as for the pion case.
   In this section we still allow both photons to be virtual, and only in
the following section we will restrict ourselves to the application we 
are interested in, namely, $e^-+\pi^+\to e^-+\pi^++\gamma$.

   We split the total VCS tensor into two parts $A$ and $B$ \cite{GellMann}, 
where class $A$ contains the pole terms, possibly together with some 
appropriate piece to ensure gauge invariance, and class $B$ contains the rest,
\begin{equation}
\label{msplit}
M^{\mu\nu}=M^{\mu\nu}_A+M^{\mu\nu}_B.
\end{equation}
   We assume that the division into $A$ and $B$ was done in such a fashion
that all symmetry principles are individually satisfied by $M^{\mu\nu}_A$
and $M^{\mu\nu}_B$.
   With the above separation, $M^{\mu\nu}_B$ is by construction regular as
$q^\mu\to 0$ or $q'^\mu\to 0$.
   In fact, there is some degree of arbitrariness concerning which
contribution is included into class $A$.
   Different choices will differ by separately gauge invariant, regular
terms (see \cite{Scherer,Fearing} for more details).

   We will now discuss a few general properties of $M^{\mu\nu}_B$.
   Under Lorentz transformations $M^{\mu\nu}_B$ transforms as a proper
second-rank Lorentz tensor which can be constructed in terms of
$q^\mu$, $q'^\mu$, and $P^\mu=p^\mu+p'^\mu$.
   A complete set of independent tensors is given by
\begin{equation}
\label{lorentztensors}
g^{\mu\nu}, P^\mu P^\nu, P^\mu q^\nu, q^\mu P^\nu,
P^\mu q'^\nu, q'^\mu P^\nu, q^\mu q^\nu, q'^\mu q'^\nu,
q^\mu q'^\nu, q'^\mu q^\nu.
\end{equation}
   These tensors are multiplied by scalar functions of the invariants
available, e.g., $q^2$, $q'^2$, $q\cdot q'$, and $q\cdot P=
q'\cdot P$.
   Symmetry with respect to charge conjugation implies that the VCS tensor
is the same for both $\pi^+$ and $\pi^-$,
\begin{equation}
\label{chargeconjugation}
M^{\mu\nu}_{B,\pi^+}(p',q';p,q)=M^{\mu\nu}_{B,\pi^-}(p',q';p,q),
\end{equation}
which can be converted into a constraint involving, say, $M^{\mu\nu}_{B,\pi^+}$
only, by making use of pion crossing (see, e.g., \cite{Barton}),
\begin{equation}
\label{pioncrossing}
M^{\mu\nu}_{B,\pi^+}(p',q';p,q)=M^{\mu\nu}_{B,\pi^+}(-p,q';-p',q),
\end{equation}
yielding finally
\begin{equation}
\label{chargeconjugationcrossing}
M^{\mu\nu}_B(q,q',P)=M^{\mu\nu}_B(q,q',-P),
\end{equation}
where from now on we omit the subscript $\pi^+$ and, using four-momentum
conservation, express $M^{\mu\nu}_B$ as a function of the three independent
momenta $q$, $q'$, and $P$.
   In order to easily implement the constraints due to photon-crossing
symmetry,
\begin{equation}
\label{crossingsymmetry}
M^{\mu\nu}_B(q,q',P)=M^{\nu\mu}_B(-q',-q,P),
\end{equation}
and the combination of charge-conjugation symmetry with pion-crossing symmetry,
Eq.\ (\ref{chargeconjugationcrossing}), we choose the following 
parametrization of $M^{\mu\nu}_B$ \cite{Fearing,Tarrach}:
\begin{eqnarray}
\label{mpar}
M^{\mu\nu}_B(q,q',P)&=&A g^{\mu\nu} + B P^\mu P^\nu
+ C (P^\mu q^\nu - q'^\mu P^\nu)
+ \tilde{C} (P^\mu q^\nu + q'^\mu P^\nu)\nonumber\\
&&+ D (P^\mu q'^\nu-q^\mu P^\nu)
+\tilde{D}(P^\mu q'^\nu+q^\mu P^\nu) + E(q^\mu q^\nu +q'^\mu q'^\nu)
\nonumber\\
&&+\tilde{E}(q^\mu q^\nu-q'^\mu q'^\nu)
+F q^\mu q'^\nu + G q'^\mu q^\nu,
\end{eqnarray}
where the scalar functions have the following properties:
\begin{eqnarray}
\label{crossingeven}
f(q^2,q'^2,q\cdot q',q\cdot P)&=&+
f(q'^2,q^2,q\cdot q',-q\cdot P),\quad \mbox{for}\quad f=A,B,C,D,E,F,G,
\\
\label{crossingodd}
f(q^2,q'^2,q\cdot q',q\cdot P)&=&-
f(q'^2,q^2,q\cdot q',-q\cdot P),\quad \mbox{for}\quad f=\tilde{C},
\tilde{D},\tilde{E},
\\
\label{ceven}
f(q^2,q'^2,q\cdot q',q\cdot P)&=&+
f(q^2,q'^2,q\cdot q',-q\cdot P),\quad \mbox{for}\quad f=A,B,E,
\tilde{E},F,G,
\\
f(q^2,q'^2,q\cdot q',q\cdot P)&=&-
f(q^2,q'^2,q\cdot q',-q\cdot P),\quad \mbox{for}\quad f=C,\tilde{C},
D,\tilde{D}.
\end{eqnarray}
   Due to gauge invariance,
\begin{equation}
\label{gaugeinvariance}
q_\mu M^{\mu\nu}_B=0,\quad M^{\mu\nu}_B  q'_\nu =0,
\end{equation}
the scalar functions of Eq.\ (\ref{mpar}) are not independent, i.e., they
are related by a homogeneous set of five independent linear equations
\cite{Fearing}.
   The constraints imposed by gauge invariance can be solved order
by order in $k$, where $k$ refers to either of $q$ or $q'$.
   This was done in \cite{Fearing}, where the structure-dependent
part up to ${\cal O}(k^4)$ was parametrized in terms of 11
low-energy coefficients, based on Lorentz invariance,
gauge invariance, crossing symmetry and the discrete symmetries.
   Alternatively, a method suggested by Bardeen and Tung \cite{Bardeen} may
be applied to construct independent invariant amplitudes which are free from
both kinematic singularities and zeros.
   In \cite{Tarrach} it was pointed out that this method requires
a slight generalization when applied to the VCS case where both
photons are virtual.
   Here we will make use of the results of \cite{Tarrach}, where it was shown
that $M^{\mu\nu}$ and thus, of course, $M^{\mu\nu}_B$ can be written as
\begin{eqnarray}
\label{mtarrach}
M^{\mu\nu}_B&=& T^{\mu\nu}_1 B_1 + T_2^{\mu\nu}
\left[B_2 -q^2 q'^2 \frac{B_6}{q\cdot q'}\right]
+T^{\mu\nu}_3\left[B_3 + (q\cdot P)^2 \frac{B_6}{q\cdot q'}\right]
\nonumber\\
&&+T_4^{\mu\nu}\left[B_4-\frac{1}{2}q\cdot P (q^2+q'^2)\frac{B_6}{q\cdot q'}
\right]
+T_5^{\mu\nu}
\left[B_5 +\frac{1}{2}q\cdot P (q^2-q'^2)\frac{B_6}{q\cdot q'}\right],
\end{eqnarray}
with
\begin{eqnarray}
\label{tb}
T_1^{\mu\nu}&=&q'^\mu q^\nu -q\cdot q' g^{\mu\nu},\nonumber\\
T_2^{\mu\nu}&=&q\cdot P(P^\mu q^\nu +q'^\mu P^\nu)
-q\cdot q' P^\mu P^\nu -(q\cdot P)^2 g^{\mu\nu},\nonumber\\
T_3^{\mu\nu}&=&q^2q'^2g^{\mu\nu}+q\cdot q' q^\mu q'^\nu
-q'^2 q^\mu q^\nu -q^2 q'^\mu q'^\nu,\nonumber\\
T_4^{\mu\nu}&=&q\cdot P (q^2+q'^2)g^{\mu\nu}-q\cdot P(q^\mu q^\nu
+q'^\mu q'^\nu)-q'^2 P^\mu q^\nu - q^2 q'^\mu P^\nu
+q\cdot q'(P^\mu q'^\nu + q^\mu P^\nu),\nonumber\\
T_5^{\mu\nu}&=&q\cdot P (q^2-q'^2)g^{\mu\nu}-q\cdot P
(q^\mu q^\nu - q'^\mu q'^\nu)+q'^2 P^\mu q^\nu - q^2 q'^\mu P^\nu
-q\cdot q' (P^\mu q'^\nu - q^\mu P^\nu).
\nonumber\\
\end{eqnarray}
   The functions $B_i$ depend on the usual scalar variables and satisfy the
following properties:
\begin{eqnarray}
\label{bpropcr}
B_i(q^2,q'^2,q\cdot q',q\cdot P)&=&\pm B_i(q'^2,q^2,q\cdot q',
-q\cdot P),\quad +:i=1,2,3,5,6,\quad -:i=4,\\
\label{bpropcc}
B_i(q^2,q'^2,q\cdot q',q\cdot P)&=&\pm B_i(q^2,q'^2,q\cdot q',
-q\cdot P),\quad +:i=1,2,3,6,\quad -:i=4,5.
\end{eqnarray}
   Each element of the tensorial basis of Eq.\ (\ref{tb}) is by construction
gauge invariant.
   The basis is not ``minimal'' in the sense that the scalar functions
multiplying the tensorial structures still contain kinematical singularities.
   In \cite{Tarrach} it was shown that it is impossible to construct such
a ``minimal'' basis.
   However, when Eq.\ (\ref{mtarrach}) is multiplied out, the $1/q\cdot q'$
singularities disappear, and the result reduces to the low-energy expression
of \cite{Fearing}.

\section{Application}
   Let us now turn to the VCS contribution to the process
$e^- +\pi^+\to e^-+\pi^+ +\gamma$, where the virtual photon generated by the
leptonic transition current is space-like, $q^2<0$, and the final photon is
real, $q'^2=0$, $q'\cdot\epsilon'=0$.
   The virtual Compton scattering tensor for this situation thus reduces to
\begin{eqnarray}
\label{mvr}
M^{\mu\nu}_B&=&[q'^\mu q^\nu-q\cdot q' g^{\mu\nu}]f_1
+[q\cdot P(P^\mu q^\nu+q'^\mu P^\nu)-q\cdot q'P^\mu P^\nu
-(q\cdot P)^2 g^{\mu\nu}]f_2\nonumber\\
&&+[q\cdot P q^2 g^{\mu\nu}- q\cdot P q^\mu q^\nu
-q^2 q'^\mu P^\nu +q\cdot q' q^\mu P^\nu]f_3,
\end{eqnarray}
where the functions $f_i$ are related to the functions $B_i$ through
\begin{eqnarray}
\label{fb}
f_1(q^2,q\cdot q', q\cdot P)&=&B_1(q^2,0,q\cdot q',q\cdot P),
\nonumber\\
f_2(q^2,q\cdot q', q\cdot P)&=&B_2(q^2,0,q\cdot q',q\cdot P),
\nonumber\\
f_3(q^2,q\cdot q', q\cdot P)&=&B_4(q^2,0,q\cdot q',q\cdot P)
+B_5(q^2,0,q\cdot q',q\cdot P).
\end{eqnarray}
   Note that for the case of at least one real photon, the terms of
Eq.\ (\ref{mtarrach}) proportional to $B_6/q\cdot q'$ precisely cancel.

   Here, we are not interested in the Bethe-Heitler contribution, where the
real photon is radiated off the initial or final electron.
   Due to current conservation at the leptonic vertex, the polarization
vector of the virtual photon can be written as
$\epsilon_\mu=e\bar{u}\gamma_\mu u/q^2$, where $u$ and $\bar{u}$ refer to
the Dirac spinors of the initial and final electron, respectively.
   We describe the reaction in the photon-pion center-of-mass system,
$\vec{p}=-\vec{q}$ and $\vec{p}\,'=-\vec{q}\,'$, and we choose the
three-momentum transfer of the initial photon to be along the $z$ axis,
$\vec{q}=\mid\vec{q}\mid\hat{e}_z$.
   All kinematical quantities can be expressed in terms of
$\omega'=\mid\vec{q}\,'\mid$, $\bar{q}\equiv\mid\vec{q}\mid$,
and $z\equiv\cos(\theta)=\hat{q}\cdot\hat{q}'$.
   Using gauge invariance of the hadronic VCS tensor, the invariant
amplitude of Eq.\ (\ref{m}) can be rewritten as
\begin{eqnarray}
\label{mgiht}
{\cal M}&=&2Mi e^2\left(\vec{\epsilon}_T\cdot\vec{M}_T +
\frac{q^2}{q^2_{0}}\epsilon_z M_z\right).
\end{eqnarray}
   Choosing the Coulomb gauge for the final real photon,
$\epsilon'^\mu=(0,\vec{\epsilon}\,')$, which
implies $\vec{\epsilon}\,'\cdot\vec{q}\,'=0$, the transverse and longitudinal
parts of ${\cal M}$ can be described in terms of two functions $A_1$, $A_2$
and one function $A_9$, respectively,
\begin{eqnarray}
\label{mt}
\vec{\epsilon}_T\cdot\vec{M}_T&=&
\vec{\epsilon}_T\cdot\vec{\epsilon}\,'^\ast (A_1+zA_2)
-\hat{q}\times\vec{\epsilon}_T\cdot\hat{q}'\times
\vec{\epsilon}\,'^\ast A_2,\\
\label{ml}
\epsilon_z M_z&=&\epsilon_z \vec{\epsilon}\,'^\ast
\cdot \hat{q} A_9,
\end{eqnarray}
where we have used the convention and nomenclature of \cite{Hemmert}.

   We now contract the parametrization of Eq.\ (\ref{mvr}) with
$\epsilon_\mu$ and $\epsilon_\nu'^\ast$, make use of Eq.\ (\ref{mgiht}),
and expand the result for ${\cal{M}}_B$ up to and including terms of order
$\omega'^2$.
   In order to keep the result as transparent as possible, we do this
in two steps.
   We first expand the kinematical factors of the tensorial
basis  in terms of $\omega'$, still keeping the functions $f_i$
with their full set of arguments.
   The class $B$ contribution to the functions $A_1$, $A_2$, and $A_9$
then reads
\begin{eqnarray}
\label{exp1a1za2}
A_1+zA_2&=&-\omega'[(\omega_0+\omega')f_1
+\omega'(4M^2-4M\omega_0+\omega_0^2-z\omega_0 \bar{q})f_2
+2M\bar{q}^2 f_3]+{\cal O}(\omega'^3),\\
\label{exp1a2}
A_2&=&-\omega'\bar{q}[f_1-\omega'(4M-\omega_0+z\bar{q})f_2
+ 2 M \omega_0 f_3] +{\cal O}(\omega'^3),\\
\label{exp1a9}
A_9&=&-\omega'\{(\omega_0 + \omega') f_1
+[- 2 M \bar{q}^2 + \omega' ( 4 M^2 - \omega_0^2 -
z \omega_0 \bar{q})] f_2\}
+{\cal O}(\omega'^3),
\end{eqnarray}
where $\omega_0\equiv q_0|_{\omega'=0}=M-\sqrt{M^2+\bar{q}^2}$
corresponds to the energy of the initial virtual photon in the limit
of zero energy of the final real photon.
   In Eqs.\ (\ref{exp1a1za2}) - (\ref{exp1a9}) we already made use of the
fact that $f_3$, in an expansion in $\omega'$, is of ${\cal O}(\omega')$.
   This property results from the definition of $f_3$, Eq.\ (\ref{fb}),
in terms of $B_4$ and $B_5$ which, in accord with their charge conjugation
properties, Eq.\ (\ref{bpropcc}), are odd functions of $q\cdot P=q'\cdot P$,
and thus must start at least as $\omega'$.
   For this statement to be true it is crucial that we have already separated
the dynamical singularities in the class $A$ contribution.

   In the next step we also expand the functions $f_i$ in terms of
$\omega'$, where we can restrict ourselves to first order in $\omega'$
since the expansion of the tensorial basis has already resulted in terms
which are at least of order $\omega'$.
   The relevant expansions read
\begin{eqnarray}
f_i(q^2,q \cdot q',q \cdot P) & = & f_i(\omega_0^2 - \bar{q}^2,0,0)
+ 2 \omega' \omega_0 f_{i,1} (\omega_0^2 - \bar{q}^2,0,0)
+ \omega' ( \omega_0 - z \bar{q})
f_{i,2} (\omega_0^2 - \bar{q}^2,0,0)
\nonumber \\
& &
+ \omega' ( 2 M - \omega_0 + z \bar{q})
f_{i,3}(\omega_0^2 - \bar{q}^2,0,0)  + {\cal{O}}(\omega'^2),
\end{eqnarray}
where $f_{i,j}$ denotes the first partial derivative of $f_i$ with
respect to the $j$th argument, i.e.,
$f_{i,1} (q^2,q\cdot q',q\cdot P) =
\frac{\partial}{\partial q^2} f_i (q^2,q\cdot q',q\cdot P)$ etc.
   Our final result for the expansion of ${\cal{M}}_B$ to second order in
$\omega'$ is
\begin{eqnarray}
\label{a1za2result}
A_1+zA_2&=&-\omega'\left\{\omega_0 f_1
+\omega'[f_1+2\omega_0^2 f_{1,1}
+\omega_0(\omega_0 - z \bar{q}) f_{1,2}
+(4M^2-4M\omega_0+\omega_0^2-z\omega_0\bar{q})f_2\right.\nonumber\\
&&\left.+2M\bar{q}^2 (2 M - \omega_0 + z\bar{q})
f_{3,3}]\right\}
+{\cal O}(\omega'^3),\\
\label{a2result}
A_2&=&-\omega'\bar{q}\left\{f_1 +\omega'[ 2 \omega_0 f_{1,1}
+(\omega_0 - z\bar{q})f_{1,2}
-(4M-\omega_0+z\bar{q})f_2\right.\nonumber\\
&&\left.+2M\omega_0( 2 M - \omega_0 + z \bar{q})f_{3,3}]\right\}
+{\cal O}(\omega'^3),\\
\label{a9result}
A_9&=&-\omega'\left\{\omega_0 f_1- 2 M \bar{q}^2 f_2
+\omega'[f_1+ 2 \omega_0^2 f_{1,1}
+ \omega_0 (\omega_0 - z \bar{q})f_{1,2}\right.\nonumber\\
&&\left.+(4 M^2 - \omega_0^2 - z \omega_0 \bar{q})f_2
- 4 M \omega_0 \bar{q}^2 f_{2,1}
- 2 M \bar{q}^2( \omega_0 - z \bar{q})f_{2,2}]\right\}
+{\cal O}(\omega'^3),
\end{eqnarray}
where the arguments of the functions $f_i$ and $f_{i,j}$ are taken
to be $(\omega_0^2-\bar{q}^2,0,0)$.
   When expanding the functions $f_i$ we explicitly made use of the
consequences of charge-conjugation symmetry, namely, $f_1$ and
$f_2$ are even functions of $q\cdot P$ and $f_3$ is odd
which follows from Eqs.\ (\ref{bpropcc}) and (\ref{fb}).

\section{Discussion}
   Eqs.\ (\ref{a1za2result}) - (\ref{a9result}) contain the central result 
of this work and serve as the starting point for discussing various
low-energy approximations.
   To be specific, we will consider the multipole expansion of 
\cite{Guichon}, comment on the limit of real Compton scattering
and, finally, compare the result of a $1/M$ expansion with the parametrization
of \cite{Fearing}.
   In order to fully appreciate the different expansion schemes it is useful
to first discuss the kinematics of $e^-+\pi^+\to e^-+\pi^++\gamma$ in
the $\omega'$-$\bar{q}$-plane.  

\subsection{Kinematical considerations}
\begin{figure}[h]
\centerline{
\epsfxsize=11.5cm
\epsfbox{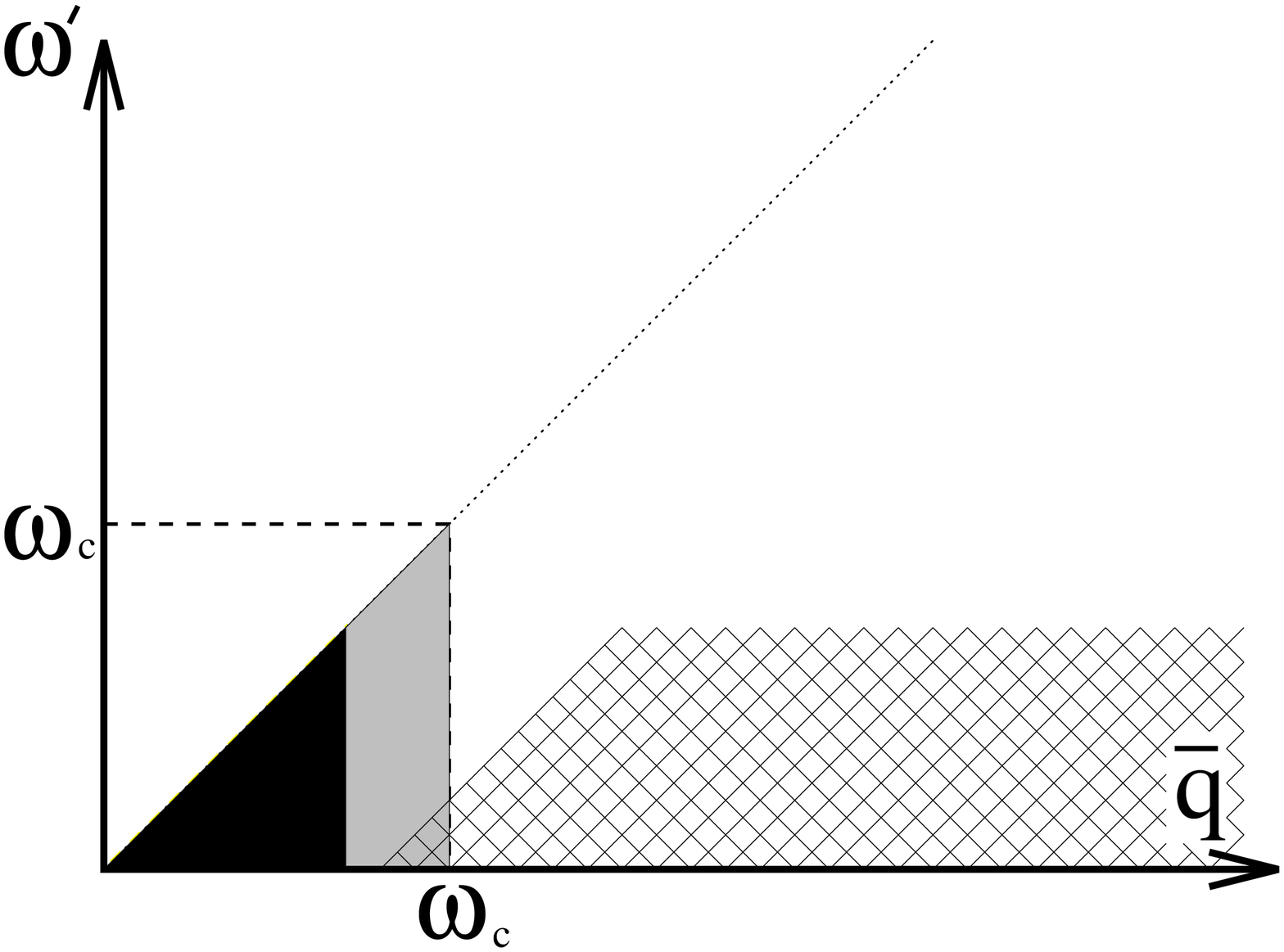}}
\caption{The $\omega'$-$\bar{q}$-plane for virtual Compton scattering
with electron-scattering kinematics ($q^2<0$, $q'^2=0$). \label{planefig}}
\end{figure}

   Fig.\ \ref{planefig} shows that region of the $\omega'$-$\bar{q}$-plane
which is accessible to electron-scattering kinematics. 
   Using energy conservation in the center-of-mass frame
and $|\omega|<\bar{q}$, one obtains
\begin{equation}
\label{energy}
\omega'+\sqrt{M^2+\omega'^2}=\omega+\sqrt{M^2+\bar{q}^2}
<\bar{q}+\sqrt{M^2+\bar{q}^2},
\end{equation}
and thus $\omega'<\bar{q}$.
   The diagonal $\omega' = \bar{q}$ corresponds to the case of real Compton
scattering.

   Let us first consider a low-energy expansion in terms of $\omega'$ 
{\em and} $\bar{q}$ as simultaneous expansion parameters which, for
example, would be a natural expansion scheme in the framework of chiral 
perturbation theory.
   In general, such an expansion is applied when $\omega'$
and $\bar{q}$ are smaller than a characteristic energy $\omega_c$ of
the model or theory in question. 
   This characteristic energy is associated with either the energy 
gap to the first particle-production threshold or the
excitation energy of the lowest excited state above the ground
state and, thus, sets an upper limit to the convergence radius of the 
low-energy expansion. 
   For example, in VCS off the nucleon $\omega_c$ is equal to the
pion mass $m_{\pi}$.
   In Fig.\ \ref{planefig} the grey area denotes the region of the
$\omega'$-$\bar{q}$-plane where such a low-energy expansion is expected
to converge.
   Clearly, if the expansion is truncated at a certain
order, the domain where it is expected to give a 
reasonable description of the full amplitude is smaller. 
   This regime is symbolically indicated by the black area of
Fig.\ \ref{planefig}.

   The multipole expansion of \cite{Guichon} is restricted to first order
in the energy of the real photon which implies that $\omega'$ has to
be small compared with $\omega_c$ but, in principle, no restrictions apply 
to $\bar{q}$.
   In particular, it is expected to work for large $\bar{q}$.
   However, when $\bar{q}$ is of the same order of magnitude as $\omega'$, 
this scheme cannot be expected to provide an adequate parametrization of the 
VCS amplitude, because terms beyond the 
linear order in $\omega'$ are likely to be equally important as the 
higher-order terms in $\bar{q}$ included in the multipole expansion. 
   This can be seen, e.g., for the term proportional to
$f_1$ in Eq.\ (\ref{a1za2result}), as soon as $\omega'$ is of the same
order as the absolute value of $\omega_0$.
   In Fig.\ \ref{planefig} the cross-hatched area schematically denotes
the domain of application of the expansion of Guichon {\em et al}.
   However, one has to keep in mind that it is difficult to decide which 
value of $\bar{q}$ is sufficiently large without an explicit model
calculation.

\subsection{Multipole expansion and generalized polarizabilities}
\label{multexp}
   We now turn to a comparison of Eqs.\ (\ref{a1za2result}) - (\ref{a9result})
with the corresponding low-energy expansion in terms of generalized 
polarizabilities as introduced by Guichon {\em et al.} \cite{Guichon}.
   These authors truncated the expansion at first order in $\omega'$:
\footnote{For details about the notation and the
definition of the generalized polarizabilities we refer the reader to
\cite{Guichon}.}
\begin{eqnarray}  \label{a1za2guichon}
A_{1}+zA_{2}&=&\omega' \sqrt{\frac{E}{M}}\biggl[
 -\sqrt{\frac{3}{2}} \omega_{0} P^{(01,01)0}(\bar{q})
 -\frac{3}{2} \bar{q}^{2} \hat{P}^{(01,1)0}(\bar{q})\biggr]
 +{\cal O}(\omega'^{2}),\\
 \label{a2guichon}
A_{2}&=&\omega' \sqrt{\frac{E}{M}} \sqrt{\frac{3}{8}}
 \bar{q} P^{(11,11)0}(\bar{q}) + {\cal O}(\omega'^{2}),\\
 \label{a9guichon}
A_{9}&=&-\omega' \sqrt{\frac{E}{M}} \sqrt{\frac{3}{2}}
 \omega_{0} P^{(01,01)0}(\bar{q}) + {\cal O}(\omega'^{2}),
\end{eqnarray}
where $E$ denotes the energy of the initial pion.
   Up to normalization factors, the quantities $P^{(01,01)0}$ and 
$P^{(11,11)0}$ are generalizations of the electric and magnetic 
polarizabilities of real 
Compton scattering (see, e.g., \cite{Lvov})
to the virtual photon case:
\begin{equation}  \label{polzsh}
\alpha(\bar{q}) = -\frac{e^{2}}{4\pi} \sqrt{\frac{3}{2}}
 P^{(01,01)0}(\bar{q}),\quad
\beta(\bar{q}) = -\frac{e^{2}}{4\pi} \sqrt{\frac{3}{8}}
 P^{(11,11)0}(\bar{q}).
\end{equation}
  The third scalar polarizability $\hat{P}^{(01,1)0}$ expresses,
to lowest order in $\omega'$,
the difference between the charge multipole and the electric multipole.

   Comparing the two low-energy expansions of Eqs.\ (\ref{a1za2result}) - 
(\ref{a9result}) and (\ref{a1za2guichon}) - (\ref{a9guichon}), we obtain
the relations 
\begin{eqnarray}  \label{alphaf}
\alpha(\bar{q})&=&\frac{e^{2}}{4\pi} \sqrt{\frac{M}{E}}
 \biggl[-f_{1}(\omega_{0}^{2}-\bar{q}^{2},0,0) +
        2M\frac{\bar{q}^{2}}{\omega_{0}}
        f_{2}(\omega_{0}^{2}-\bar{q}^{2},0,0) \biggr], \\
 \label{betaf}
\beta(\bar{q})&=&\frac{e^{2}}{4\pi} \sqrt{\frac{M}{E}}
 f_{1}(\omega_{0}^{2}-\bar{q}^{2},0,0), \\
 \label{pmixf}
\frac{e^{2}}{4\pi} \hat{P}^{(01,1)0}(\bar{q})&=&
 \frac{e^{2}}{4\pi} \sqrt{\frac{M}{E}} \frac{4}{3} M
 f_{2}(\omega_{0}^{2}-\bar{q}^{2},0,0).
\end{eqnarray}
   From Eqs.\ (\ref{alphaf}) - (\ref{pmixf}), it is now evident that one of the
three polarizabilities may be written as a linear combination of the remaining
two. 
   For instance, we can eliminate $\hat{P}^{(01,1)0}$ in favor of 
$\alpha(\bar{q})$ and $\beta(\bar{q})$,
\begin{equation}  \label{pmixab}
\frac{e^{2}}{4\pi} \hat{P}^{(01,1)0}(\bar{q})=
 \frac{2\omega_{0}}{3\bar{q}^{2}} \big[\alpha(\bar{q})+\beta(\bar{q})\big],
\end{equation}
which is exactly the relation that has been found within the framework of the 
linear $\sigma$ model \cite{Metz}.
   We stress that this result is due to the constraint of 
Eq.\ (\ref{chargeconjugationcrossing}), and therefore ultimately follows 
from the symmetry with respect to charge conjugation and pion crossing.
   In the multipole expansion of \cite{Guichon} no use has been made of this 
symmetry.
   To be specific, without this constraint the function $f_{3}$ would appear
in the transverse amplitudes, Eqs.\ (\ref{a1za2result}) and (\ref{a2result}),
already at linear order in $\omega'$, as can be seen from 
Eqs.\ (\ref{exp1a1za2}) and (\ref{exp1a2}), resulting in 
one additional independent function. 
   In the framework of \cite{Fearing}, this would correspond to the term
proportional to $\tilde{e}_1$, indicating a violation of charge-conjugation
or time-reversal symmetry.
   
   The most surprising consequence of Eq.\ (\ref{pmixab}) concerns the 
low-energy behavior of the spin-independent electric multipole
$H^{(21,21)0}(\omega',\bar{q})$, describing electric dipole radiation
in both the initial and final states.
   Using Eq.\ (\ref{pmixab}) one gets
\begin{equation}  \label{multipole}
H^{(21,21)0}(\omega',\bar{q})=\frac{4\pi}{e^{2}} \sqrt{\frac{8}{3}}
 \omega' \omega_{0} \beta(\bar{q}) + {\cal O}(\omega'^{2}),
\end{equation}
i.e., to lowest order in $\omega'$ the electric multipole, for all
$\bar{q}$, is given by the generalized magnetic polarizability.
   Since $\omega_{0}\approx -\bar{q}^2/2M$, the right-hand side
of Eq.\ (\ref{multipole}) vanishes in the static limit, $M \to \infty$.
   Therefore the relation between $H^{(21,21)0}$  and $\beta$ is a recoil
effect and not due to an intrinsic property of the target.
   Nevertheless, it is interesting to note that the magnetic polarizability 
determines the recoil contribution of an electric multipole, 
even though, after all, it might not be so surprising, since it is well-known 
that electric and magnetic effects mix when transforming from one frame to 
another.

   Finally, we emphasize that, as a result of Eq.\ (\ref{pmixab}),
to lowest order in $\omega'$ both transverse amplitudes,
Eqs.\ (\ref{a1za2result}) and (\ref{a2result}), are completely given in terms 
of the magnetic polarizability.
   The electric polarizability $\alpha$, as defined in Eq.\ (\ref{alphaf}),
is part of the $\omega'^{2}$ contribution to the amplitude $A_{1}+zA_{2}$,
which can be seen by making use of the identity 
$\bar{q}^{2} = \omega_{0}^{2} - 2M\omega_{0}$.
   However, since at the same order there are other independent
contributions in Eq.\ (\ref{a1za2result}), $\alpha(\bar{q})$ cannot
be determined from this amplitude.
   Thus, contrary to real Compton scattering, in VCS it is impossible to
extract the generalized electric polarizability from the transverse amplitude,
and one has to resort to the longitudinal amplitude $A_{9}$ in order to obtain 
$\alpha(\bar{q})$.

\subsection{Real Compton scattering}
   We now take the limit of real Compton scattering (RCS) in 
Eqs.\ (\ref{a1za2result}) and (\ref{a2result}),
$\omega\equiv\bar{q}=\omega'$, considering terms up to second order
in $\omega$.
   Of course, the contribution of the longitudinal amplitude to the invariant
matrix element vanishes. 
   Making use of the expansion 
$\omega_{0}=-\omega^{2}/2M+{\cal O}(\omega^{4})$, we 
obtain 
\begin{eqnarray} \label{a1za2RCS}
A_{1}^{RCS}+zA_{2}^{RCS}&=&-\omega^{2}
  \left[f_{1}(0,0,0)+4M^{2} f_{2}(0,0,0)\right]+{\cal O}(\omega^{4})
  =\frac{4\pi}{e^{2}}\omega^{2}\alpha(0)+{\cal O}(\omega^{4}), \\
 \label{a2RCS}
A_{2}^{RCS}&=&-\omega^{2}f_{1}(0,0,0)+{\cal O}(\omega^{4})
  =-\frac{4\pi}{e^{2}}\omega^{2}\beta(0)+{\cal O}(\omega^{4}),
\end{eqnarray}
leading to the correct low-energy behavior of the RCS amplitudes
\cite{Lvov}.
   We stress that in order to obtain this result, it is mandatory to keep the 
terms quadratic in $\omega'$ in Eqs.\ (\ref{a1za2result}) and
(\ref{a2result}).
   These terms are beyond the accuracy of the multipole expansion of 
\cite{Guichon}.

\subsection{Low-energy expansion}
   In \cite{Fearing} the structure-dependent class-$B$ contribution was
parametrized up to and including terms of fourth order in $q$ and $q'$.
   Recently, the corresponding structure coefficients for VCS off the nucleon 
have been calculated within the framework of heavy-baryon chiral perturbation 
theory to third order in the momenta \cite{Hemmert}, by expanding
the invariant amplitude in terms of $\omega'$ and $\bar{q}$ 
simultaneously.
   Our general expansion in Eqs.\ (\ref{a1za2result}) - (\ref{a9result}) can
be compared with a heavy-baryon calculation if we expand 
Eqs.\ (\ref{a1za2result}) - (\ref{a9result}) 
in terms of $\bar{q}^{2}$ and neglect all but the leading terms of a 
$1/M$ expansion.
   In our final result we only list the terms to quadratic
order in $\omega'$ and to quartic order in $r$, where
$r \in \{\omega',\bar{q}\}$:
\begin{eqnarray} \label{a1za2HB}
A_{1}^{HB} + zA_{2}^{HB} &=& -\omega'^{2}\Big\{f_{1}(0,0,0) +
  4M^{2}f_{2}(0,0,0) \nonumber\\
 & & \quad - \bar{q}^{2}\left[f_{1,1}(0,0,0) + 4M^{2}f_{2,1}(0,0,0)
  - 4M^{2}f_{3,3}(0,0,0)\right]\Big\} + {\cal O}(\omega'^{3}), \\
 \label{a2HB}
A_{2}^{HB} &=& -\omega'\bar{q}\Big\{f_{1}(0,0,0) - \bar{q}^{2}f_{1,1}(0,0,0)
  - \omega'\bar{q}zf_{1,2}(0,0,0)\Big\} + {\cal O}(\omega'^{3}), \\
 \label{a9HB}
A_{9}^{HB} &=& -\omega'^{2}\Big\{f_{1}(0,0,0) + 4M^{2}f_{2}(0,0,0)\nonumber \\
 & & \quad - \bar{q}^{2}\left[f_{1,1}(0,0,0) +
  4M^{2}f_{2,1}(0,0,0)\right]\Big\} + {\cal{O}}(\omega'^{3}). 
\end{eqnarray}
   We find the following identities for the structure constants defined
in \cite{Fearing}:
\begin{eqnarray} \label{structurerelation}
g_0 & = & f_1(0,0,0),\nonumber\\
{\tilde{c}}_1 & = & \frac{1}{2} f_2(0,0,0), \nonumber\\
g_{2a} & = & f_{1,2}(0,0,0), \nonumber\\
g_{2b} & = & f_{1,1}(0,0,0), \nonumber\\
c_3 & = & \frac{1}{4} f_{3,3}(0,0,0), \nonumber\\
{\tilde{c}}_{3b} & = & \frac{1}{2} f_{2,1}(0,0,0)
 - \frac{1}{4} f_{3,3}(0,0,0).
\end{eqnarray}
   The remaining three structure constants of \cite{Fearing} involve
terms of ${\cal{O}}(\omega'^3 \bar{q})$ and ${\cal{O}}(\omega'^4)$
and, thus, cannot be related to the functions $f_i$ by means of
Eqs.\ (\ref{a1za2result}) - (\ref{a9result}).
   Furthermore, the presence of the $f_{3,3}$ piece in Eq.\ (\ref{a1za2HB})
makes it impossible to extract the derivative 
$\frac{d}{d\bar{q}^{2}} \alpha (\bar{q} = 0 )$ from the
$\omega'^2 \bar{q}^{2}$ term.
   However, in the longitudinal part of the amplitude, the $f_{3,3}$
piece is absent and the coefficients of the $\omega'^2 \bar{q}^{2}$ term
add up to the slope of the electric polarizability with respect to
$\bar{q}^{2}$ (see the discussion at the end of subsection \ref{multexp}).

\section{Summary and conclusion}
   We discussed the general amplitude for VCS off a spinless target.
   The results may also be applied to the spin-averaged amplitude of
the nucleon case.
   We restricted our considerations to the matrix element involving a 
spacelike virtual photon in the initial state and a real photon in the final 
state which  can be expressed in terms of one longitudinal and two transverse 
amplitudes.
   We assumed that the general matrix element may be separated into
a pole contribution and a residual part which is regular as either
of the two photon four-momenta approaches zero.
   We then discussed a low-energy expansion of the regular amplitude up
to and including terms of second order in the frequency $\omega'$ of the final
photon, without restrictions on the absolute value $\bar{q}$ of 
the three-momentum of the initial virtual photon.
   A multipole expansion, truncated at first order in the energy of the final 
photon, results in two independent functions (generalized polarizabilities) 
instead of three as previously claimed.
   This reduction is obtained as a consequence of charge-conjugation 
invariance in combination with pion (or nucleon) crossing.
   Whether charge-conjugation symmetry also leads to a reduction in the 
number of spin-dependent generalized polarizabilities remains to be seen. 
   At leading order in $\omega'$, we found that both transverse amplitudes 
are determined by $\beta(\bar{q})$, the generalization of the magnetic
polarizability of RCS to arbitrary $\bar{q}$.
   On the other hand, the generalized electric polarizability 
$\alpha(\bar{q})$ appears in the longitudinal amplitude only.
   Even in an expansion to second order in $\omega'$, the generalized
electric polarizability cannot be extracted from the transverse part 
since additional independent terms appear at the same order.
   Furthermore, at leading order the (E1,E1) transition matrix element is 
governed by the generalized magnetic polarizability and vanishes in the 
static limit, indicating a recoil effect.
   In order to obtain the standard limit of RCS 
involving the usual electromagnetic polarizabilities $\alpha(0)$
and $\beta(0)$, it is necessary to include the terms
of second order in $\omega'$, being so far beyond the standard analysis 
of VCS in terms of generalized polarizabilities.
   Finally, we performed a $1/M$ expansion as used in a heavy-baryon 
calculation and, within that framework, established the connection 
between the general expression and the coefficients of a recently proposed 
low-energy expansion.

\section{Acknowledgements}
This work was supported by the Deutsche Forschungsgemeinschaft (SFB 201).
A.\ M.\ would like to thank P.\ A.\ M.\ Guichon for a stimulating
discussion.

\end{document}